\documentclass[11pt]{article} 
\usepackage[top=20mm,bottom=20mm, left=20mm, right=20mm]{geometry}

\usepackage{authblk}
\usepackage{enumerate}
\usepackage{amsmath, cite, mathrsfs}
\usepackage{graphicx, color, hyperref}
\usepackage{amssymb, amsthm, makecell}
\usepackage{stmaryrd}

 \hypersetup{
     colorlinks=true,       
     linkcolor=red,          
     citecolor=blue,        
     filecolor=magenta,      
     urlcolor=cyan           
 }

\newtheorem{lemma}{Lemma}[section]

\newtheorem{ids}{\bf Identities}[section]

\def\div{\operatorname{div}}
\def\Div{\operatorname{Div}}

\def\curl{\operatorname{curl}}
\def\Curl{\operatorname{Curl}}
\def\sym{\operatorname{sym}}

\def\tr{\operatorname{tr}}
\def\Lin{\operatorname{Lin}}
\def\Sym{\operatorname{Sym}}

\def\da{\operatorname{da}}
\def\dl{\operatorname{dl}}

\def\dv{\operatorname{dv}}

\title{Some Consequences of the Distributional Stress Equilibrium Condition}
\author{Animesh Pandey and Anurag Gupta\thanks{ag@iitk.ac.in}}
\date{Department of Mechanical Engineering, Indian Institute of Technology Kanpur, 208016, India\\[2ex]%
    \today
}

\begin{document}
\maketitle

\begin{abstract}
We derive two consequences of  the distributional form of the stress equilibrium condition while incorporating piecewise smooth stress and body force fields with singular concentrations on an interface. First we obtain the local equilibrium conditions in the bulk and at the interface, the latter including conditions on the interfacial stress and stress dipole. Second we obtain the necessary and the sufficient conditions on the divergence-free non-smooth stress field for there to exist a stress function field such that the equilibrium is trivially satisfied. In doing so we allow the domain to be non-contractible with 
mutually disjoint connected boundary components. Both derivations illustrate the utility of the theory of distributions in dealing with singular stress fields.

\vspace{4mm}\noindent \textbf{Mathematics Subject Classification.} 74A10, 74G70

\vspace{4mm}\noindent \textbf{Keywords.} Distributional stress equilibrium; Singular stress fields; Stress concentrations; Stress function for non-contractible domains; Stress dipole; Force dipole wall.

\end{abstract}

\section{Introduction} \label{intro}

Let $\Omega \subset \mathbb{R}^3$ be a bounded, connected, open set, with a smooth boundary $\partial \Omega$. Let $S \subset \Omega$ be a regular oriented surface of bounded area, with unit normal $\boldsymbol{n}$ and boundary $\partial S$, such that $S$ is either a closed surface or its boundary is completely contained within the boundary of $\Omega$ (i.e., $\partial S - \partial \Omega = \emptyset$). We recall two familiar results~\cite{gurtin1973linear, gurtin1975continuum}. (i) Let $\boldsymbol{\sigma}$ be the piecewise smooth bulk Cauchy stress on $\Omega$, possibly discontinuous across $S$, and $\boldsymbol{\sigma}_1$ be the smooth interfacial Cauchy stress on $S$. The interfacial stress can be interpreted as the excess stress which remains unaccounted in a purely bulk consideration of the theory. Let $\boldsymbol{b}$  be the piecewise smooth bulk body force field on $\Omega$ and let $\boldsymbol{b}_1$ be the smooth body force field concentrated on $S$.  The stress equilibrium equations in the interior of $\Omega$ are then given as $\div \boldsymbol{\sigma} + \boldsymbol{b} = \boldsymbol{0}$, for all $\boldsymbol{x} \in \Omega - S$, and $\div_S \boldsymbol{\sigma}_1 - \llbracket \boldsymbol{\sigma} \rrbracket \boldsymbol{n} + \boldsymbol{b}_1= \boldsymbol{0}$, for all $\boldsymbol{x} \in S$, with $\boldsymbol{\sigma}_1 \boldsymbol{n} = \boldsymbol{0}$. In these expressions $\div_S$ represents the surface divergence (analogously, $\nabla_S$ will represent the surface gradient) and $\llbracket \boldsymbol{\sigma} \rrbracket (\boldsymbol{x}) = \boldsymbol{\sigma}^+(\boldsymbol{x}) -  \boldsymbol{\sigma}^-(\boldsymbol{x})$, where $\boldsymbol{\sigma}^\pm(\boldsymbol{x})$ are limiting values of $\boldsymbol{\sigma}$ as $\boldsymbol{x}$ is approached from the two sides of $S$ (the negative side is the one into which $\boldsymbol{n}$ points). (ii) Let $\boldsymbol{\sigma}$ represent the smooth bulk Cauchy stress on $\Omega$. Let $S=\emptyset$ and $\boldsymbol{b} = \boldsymbol{0}$. Then, for a $\Omega$ whose boundary $\partial \Omega$ is a single connected component, $\div \boldsymbol{\sigma} = \boldsymbol{0}$ is equivalent to the existence of a smooth stress function field $\boldsymbol{\phi}$ such that $\curl \curl  \boldsymbol{\phi} = \boldsymbol{\sigma}$.

In this note we use a recently developed framework~\cite{pandey2020topological, PANDEY2021103806}, within which several singular problems in elasticity were resolved, to obtain novel generalizations of the two results mentioned above.  The framework is based on the theory of distributions~\cite{friedlander1998introduction}. The relevant mathematical preliminaries are summarized in Section~\ref{prelim}; for details we refer the reader to our recent work~\cite{pandey2020topological, PANDEY2021103806}. The starting point for the proposed generalization is the distributional stress equilibrium condition, see Section~\ref{stressequibsec}. The posited equilibrium condition holds for singular stress fields. It remains meaningful even when the Cauchy flux maps fail to exist for all smooth surfaces in $\Omega$. We use it, in Section~\ref{concent}, to derive local stress equilibrium conditions in the interior of $\Omega$ when the stress field consists of a bulk part, an interfacial concentration, and an interfacial concentration of stress dipole. The stress dipole distribution on $S$ can be interpreted in terms of two neighbouring surfaces of stress concentration with opposite sign. Physically, it can appear in response to a concentration of body force dipoles (or quadrupoles) over $S$ which in turn could originate due to a defect concentration at the surface. Next, in Section~\ref{noncon}, we establish the existence of a (distributional) stress function field, which yields an equilibrated stress, when the boundary of the domain $\Omega$ has mutually disjoint, albeit connected, components (for instance, a hollow sphere or a hollow torus). Towards this end, we use a consequence of  the De Rham's theorem (see Lemma~\ref{LemmaNSCurlFields}) in order to obtain the necessary and sufficient conditions on a distributional vector field for it to be expressible as a curl of another distributional vector field. We obtain explicit conditions on stress for the case where it has a piecewise smooth bulk part in addition to singular concentrations on an interface. Our result extends an earlier work by Gurtin~\cite{gurtin1963generalization} which was restricted to smooth stress fields without any concentrations. 

\noindent \textit{Notation}. We denote the spaces of vectors, tensors, and symmetric tensors by $\mathcal{V}$, $\Lin$, and $\Sym$, respectively. The identity tensor in $\Lin$ is denoted by $\boldsymbol{I}$. The inner product, cross product, and the dyadic product are represented by $\langle \cdot , \cdot \rangle$, $\times$, and $\otimes$, respectively. The trace of a tensor $\boldsymbol{A} \in \Lin$ is defined as $\tr \boldsymbol{A}= \langle \boldsymbol{A},\boldsymbol{I} \rangle$. 
 We use $C^0 (\Omega)$, $C^{\infty}(\Omega)$, and $C^{r}(\Omega)$ ($r$ is a positive integer), to represent spaces of continuous, smooth, and $r$-times differentiable functions on $\Omega$, respectively. For an open set $\Omega \subset \mathbb{R}^3$ let $\mathcal{D}(\Omega)$, $\mathcal{D}(\Omega,\mathcal{V})$, and $\mathcal{D}(\Omega,\Lin)$ represent the spaces of compactly supported smooth functions from $\Omega$ to $\mathbb{R},$ $\mathcal{V}$, and $\Lin$, respectively. The gradient, divergence, and curl operators on smooth fields are denoted by $\nabla$, $\div$, and $\curl$, respectively.

\section{Some results from the distribution theory} \label{prelim}

The spaces of scalar, vector, and tensor valued distributions, represented by $\mathcal{D}'(\Omega)$, $\mathcal{D}'(\Omega,\mathcal{V})$, and $\mathcal{D}'(\Omega,\Lin)$, respectively, are dual to the respective spaces of compactly supported smooth functions on $\Omega$. 
The divergence of a vector valued distribution $\boldsymbol{T}\in \mathcal{D}'(\Omega,\mathcal{V})$ is a scalar valued distribution defined as $\Div \boldsymbol{T}(\psi)=-\boldsymbol{T}(\nabla \psi)$, for all $\phi \in \mathcal{D}(\Omega)$. The divergence of a tensor valued distribution $\boldsymbol{T}\in \mathcal{D}'(\Omega,\Lin)$ is a vector valued distribution defined as $\Div \boldsymbol{T}(\boldsymbol{\psi})=-\boldsymbol{T}(\nabla \boldsymbol{\psi})$, for all $\boldsymbol{\psi} \in \mathcal{D}(\Omega,\mathcal{V})$. The curl of a vector valued distribution $\boldsymbol{T}\in \mathcal{D}'(\Omega,\mathcal{V})$ is a vector valued distribution defined as $\Curl \boldsymbol{T}(\boldsymbol{\psi})=\boldsymbol{T}(\curl \boldsymbol{\psi})$, for all $\boldsymbol{\psi} \in \mathcal{D}(\Omega,\mathcal{V})$. The curl of a tensor valued distribution $\boldsymbol{T}\in \mathcal{D}'(\Omega,\Lin)$ is a tensor valued distribution defined as $\Curl \boldsymbol{T}(\boldsymbol{\psi}^T)=\boldsymbol{T}((\curl \boldsymbol{\psi})^T)$, for all $\boldsymbol{\psi} \in \mathcal{D}(\Omega,\Lin)$, where the superscript $T$ denotes the transpose. A sequence of distributions $T_j \in \mathcal{D}'(\Omega)$ is said to converge to $T_0 \in \mathcal{D}'(\Omega)$, as $j\to \infty$, in the sense of distributions, that is $T_j \to T_0,$ if $T_j (\phi) \to T_0 (\phi)$, for all $\phi \in \mathcal{D}(\Omega)$. 

Consider a smooth vector field $\boldsymbol{v}\in C^\infty(\Omega,\mathcal{V})$ and a domain $\Omega \subset \mathbb{R}^3$ whose boundary has only one disjoint connected component. Then, the necessary and sufficient conditions for the existence of $\boldsymbol{\phi}\in C^\infty(\Omega,\mathcal{V})$, satisfying $\curl \boldsymbol{\phi}=\boldsymbol{v}$, is $\div \boldsymbol{v}=0$. The following lemma extends this result for distributional vector fields and  domains whose boundary can have multiple mutually disjoint connected components. It was established in our recent work~\cite{PANDEY2021103806} as a consequence of De Rham's theorem.
\begin{lemma}
\label{LemmaNSCurlFields}
Given a vector valued distribution $\boldsymbol{P}\in \mathcal{D}'(\Omega,\mathcal{V})$ there exists $\boldsymbol{Q} \in \mathcal{D}'(\Omega,\mathcal{V})$ such that $\Curl \boldsymbol{Q} = \boldsymbol{P}$ if and only if $\boldsymbol{P}(\boldsymbol{\phi})=0$ for all $\boldsymbol{\phi}\in \mathcal{D}(\Omega,\mathcal{V})$ such that $\curl \boldsymbol{\phi}=\boldsymbol{0}$.
\end{lemma}
\noindent For every curl free compactly supported smooth vector field $\boldsymbol{\phi}\in \mathcal{D}(\Omega,\mathcal{V})$, such as those which appear in the above lemma, there exists a smooth scalar field $u \in C^{\infty}(\Omega)$ which satisfies $\nabla u = \boldsymbol{\phi}$ in $\Omega$ and $u=c_i$ on $\partial \Omega_i$, where $c_i \in \mathbb{R}$ are constants with $c_0=0$. 

In this note we will be interested in three specific types of distributions. Let $\mathcal{B}(\Omega,\mathcal{V}) \subset \mathcal{D}' (\Omega,\mathcal{V})$ be a family of vector valued distributions such that for every $\boldsymbol{B} \in \mathcal{B}(\Omega,\mathcal{V})$ there exists a piecewise smooth vector valued function $\boldsymbol{b}$ on $\Omega$, possibly discontinuous across $S$ (with $\partial S - \partial \Omega = \emptyset$), satisfying
\begin{equation}
\boldsymbol{B}(\boldsymbol{\phi})=\int_{\Omega} \langle \boldsymbol{b} , \boldsymbol{\phi} \rangle \dv, \label{distB}
\end{equation}
for all $\boldsymbol{\phi} \in \mathcal{D}(\Omega,\mathcal{V})$, where $\dv$ is the volume measure on $\Omega$. The discontinuity in $\boldsymbol{b}$ is assumed to be a smooth function on $S$. 
Let $\mathcal{C}(\Omega,\mathcal{V}) \subset \mathcal{D}' (\Omega,\mathcal{V})$ be a family of vector valued distributions such that for every $\boldsymbol{C} \in \mathcal{C}(\Omega,\mathcal{V})$ there exists a smooth vector valued function $\boldsymbol{c}$ on $S$ satisfying
\begin{equation}
\boldsymbol{C} (\boldsymbol{\phi})=\int_{S} \langle \boldsymbol{c}, \boldsymbol{\phi} \rangle \da, \label{distC}
\end{equation}
for all $\boldsymbol{\phi} \in \mathcal{D}(\Omega,\mathcal{V})$, where $\da$ is the area measure on the surface. 
Let $\mathcal{F}(\Omega,\mathcal{V}) \subset \mathcal{D}' (\Omega,\mathcal{V})$ be a family of vector valued distributions such that for every $\boldsymbol{F} \in \mathcal{F}(\Omega,\mathcal{V})$ there exists a vector valued smooth function $\boldsymbol{f}$ on $S$ satisfying
\begin{equation}
\boldsymbol{F} (\boldsymbol{\phi})=\int_{S} \left\langle \boldsymbol{f}, \frac{\partial \boldsymbol{\phi}}{\partial {n}} \right\rangle \da, \label{distF}
\end{equation}
for all $\boldsymbol{\phi} \in \mathcal{D}(\Omega,\mathcal{V})$, where ${\partial}/{\partial {n}}$ represents the partial derivative along $\boldsymbol{n}$, i.e., ${\partial \boldsymbol{\phi}}/{\partial n}= (\nabla \boldsymbol{\phi})\boldsymbol{n}$. The respective tensor valued distributional spaces, for instance $\mathcal{B}(\Omega,\Lin)$, can be introduced analogously. The following sets of identities, taken from our previous work~\cite{pandey2020topological, PANDEY2021103806}, will be useful in later sections. 
\begin{ids}
\textup{\cite{pandey2020topological}}
\label{DivergenceLemma} (Divergence of distributions) Let the open set $\Omega  \subset \mathbb{R}^3$ be such that $\partial S - \partial \Omega = \emptyset$. 
For ${\psi} \in \mathcal{D}(\Omega)$,

\noindent (a) If $\boldsymbol{B} \in \mathcal{B}(\Omega,\mathcal{V})$, as defined in \eqref{distB}, then
\begin{equation}
\begin{split}
\Div \boldsymbol{B} \left({\psi}\right)= \int_{\Omega}  (\div \boldsymbol{b}) {\psi}  \dv - \int_{S} \left\langle \llbracket \boldsymbol{b} \rrbracket , \boldsymbol{n} \right\rangle {\psi} \da. 
\end{split} \label{DivB}
\end{equation}

\noindent (b) If $\boldsymbol{C} \in  \mathcal{C}(\Omega,\mathcal{V})$, as defined in \eqref{distC}, then
\begin{equation}
\label{DivC}
\Div \boldsymbol{C} \left({\psi}\right)=  \int_{S}   \left(\div_S \boldsymbol{c} + \kappa \langle\boldsymbol{c} ,\boldsymbol{n}\rangle \right){\psi} \da -\int_S \left\langle \boldsymbol{c}, \boldsymbol{n}   \right\rangle \frac{\partial {\psi}}{\partial {n}} \da,
\end{equation}
where $\kappa$ is twice the mean curvature of surface $S$.

\noindent (c) If  $\boldsymbol{F} \in \mathcal{F}(\Omega,\mathcal{V})$, as defined in \eqref{distF}, then
\begin{equation}
\label{DivF}
\Div \boldsymbol{F}({\psi})= - \int_S \div_S \left( (\nabla_S \boldsymbol{n}) \boldsymbol{f} \right) {\psi} \da + \int_S \left(\div_S \boldsymbol{f} + \kappa \langle\boldsymbol{f} ,\boldsymbol{n}\rangle \right) \frac{\partial {\psi}}{\partial n}  \da - \int_S \langle \boldsymbol{f},\boldsymbol{n} \rangle \langle \nabla(\nabla {\psi}), \boldsymbol{n} \otimes \boldsymbol{n} \rangle \da.
\end{equation}

\end{ids}

\begin{ids}
\textup{\cite{PANDEY2021103806}}
\label{ActionOnGradientFields} 
Let the open set $\Omega  \subset \mathbb{R}^3$ has a boundary $\partial \Omega$ consisting of $k$ ($0\leq i \leq k-1$) mutually disjoint smooth connected components, each denoted as $\partial \Omega_i$. Let $\partial S - \partial \Omega = \emptyset$. For $\boldsymbol{u} \in C^{\infty}(\Omega,\mathcal{V})$ which satisfies $\boldsymbol{u}=\boldsymbol{c}_i$ on $\partial \Omega_i$, where $\boldsymbol{c}_i \in \mathcal{V}$ are constant vectors and $\boldsymbol{c}_0=\boldsymbol{0}$, such that $\nabla \boldsymbol{u} = \boldsymbol{v} \in \mathcal{D}(\Omega,\Lin)$,

\noindent (a) If $\boldsymbol{B} \in \mathcal{B}(\Omega,\Lin)$ then
\begin{equation}
\label{PiecewiseSmooth}
\begin{split}
 \boldsymbol{B} \left({\nabla \boldsymbol{u}}\right)= -\int_{\Omega}  \langle \div \boldsymbol{b}, \boldsymbol{u} \rangle  \dv + \int_{S} \left\langle \llbracket \boldsymbol{b} \rrbracket \boldsymbol{n}, \boldsymbol{u} \right\rangle \da+ \sum_{1 \leq i < k-1} \left\langle \boldsymbol{c}_i, \left(\int_{\partial \Omega_i}   \boldsymbol{b}   \boldsymbol{n} \da \right) \right\rangle . 
\end{split} 
\end{equation}

\noindent (b) If $\boldsymbol{C} \in  \mathcal{C}(\Omega,\Lin)$ then
\begin{equation}
\label{DivC}
 \boldsymbol{C} \left({\nabla \boldsymbol{u}}\right)=  -\int_{S}   \left\langle \left(\div_S \boldsymbol{c} + \kappa \boldsymbol{c} \boldsymbol{n} \right), \boldsymbol{u} \right\rangle \da +\int_S \left\langle \boldsymbol{c} \boldsymbol{n}, \frac{\partial \boldsymbol{u}}{\partial {n}}   \right\rangle \da + \sum_{1 \leq i \leq k-1} \left( \left\langle \boldsymbol{c}_i, \int_{\partial S \cap \partial \Omega_i}   \boldsymbol{c} \boldsymbol{\nu}   \dl \right)\right\rangle,
\end{equation}
where $\boldsymbol{\nu}$ is the in-plane normal to ${\partial S \cap \partial \Omega_i}$ and $\dl$ is the length measure on ${\partial S \cap \partial \Omega_i}$.

\noindent (c) If  $\boldsymbol{F} \in \mathcal{F}(\Omega,\Lin)$ then
\begin{equation}
\label{DivF}
\begin{split}
\boldsymbol{F}({\nabla \boldsymbol{u}})= \int_S \left\langle\div_S \left( (\boldsymbol{f} \nabla_S \boldsymbol{n})  \right),  \boldsymbol{u} \right\rangle \da - \int_S \left\langle \left(\div_S \boldsymbol{f} + \kappa \boldsymbol{f} \boldsymbol{n} \right), \frac{\partial \boldsymbol{u}}{\partial n} \right\rangle  \da  + \int_S \left\langle \boldsymbol{f}\boldsymbol{n},   \nabla(\nabla \boldsymbol{u}) \boldsymbol{n} \otimes \boldsymbol{n} \right\rangle \da - \\   \sum_{1 \leq i \leq k-1}  \left\langle \boldsymbol{c}_i, \int_{\partial S \cap \partial \Omega_i}  (\boldsymbol{f}\nabla_S \boldsymbol{n})  \boldsymbol{\nu} \dl \right\rangle.
\end{split}
\end{equation}
\end{ids}

\section{Distributional stress equilibrium condition} \label{stressequibsec}

Given a distributional stress field $\boldsymbol{\Sigma} \in \mathcal{D}'(\Omega,\Sym)$ and a distributional body force field $\boldsymbol{B} \in \mathcal{D}'(\Omega,\mathcal{V})$,  the distributional equilibrium condition is postulated as~\cite{pandey2020topological}
\begin{equation}
\label{DistributionalBalanceLaw}
\Div \boldsymbol{\Sigma} + \boldsymbol{B}=\boldsymbol{0}.
\end{equation}
The distributional equilibrium condition generalizes the classical stress equilibrium relation to situations when stress and body force fields are possibly singular over $\Omega$; the former reduces to the latter if the fields are smooth over $\Omega$. For singular fields, however, the equilibrium conditions may be be given in terms of pointwise fields and it may be imperative to state the equilibrium in distributional terms~\cite{pandey2022point}. In fact, the distributional stress equilibrium holds even when the Cauchy flux maps fail to exist for all surfaces in $\Omega$.  The Cauchy flux map $\mathcal{F}$ is defined as 
\begin{equation}
\label{CauchyFluxasLimit}
\mathcal{F}(\mathcal{S})=\lim_{\rho \to 0} \int_{\mathcal{S}} \boldsymbol{\sigma}_\rho \boldsymbol{n} {\da}
\end{equation}
for any smooth oriented surface $\mathcal{S}$ in $\Omega$ with unit normal $\boldsymbol{n}$, where $\boldsymbol{\sigma}_\rho$ is a sequence of smooth fields such that $\boldsymbol{\sigma}_\rho \to \boldsymbol{\sigma}$ in the sense of distributions~\cite{vsilhavy1987existence, vsilhavy2008cauchy}. 
The map $\mathcal{F}(\mathcal{S})$ represents the contact force transmitted across the {curve} $\mathcal{S}$. Given a Cauchy flux map and a smooth body force field $\boldsymbol{b}$, the equilibrium condition is given by
\begin{equation}
\label{BalanceLaw}
\mathcal{F}(\partial \mathcal{P})+\int_\mathcal{P} \boldsymbol{b} \da=\boldsymbol{0},
\end{equation}
where $\mathcal{P} \subset \Omega$ is an arbitrary open subset of $\Omega$ with smooth boundary $\partial \mathcal{P}$. If the stress field is smooth then this condition is equivalent to $\div \boldsymbol{\sigma} + \boldsymbol{b}=\boldsymbol{0}$ in $\Omega$. The symmetry of the smooth stress field is a consequence of the angular momentum balance.
For a general stress field $\boldsymbol{\sigma}\in \mathcal{D}'(\Omega,\Sym)$, however, the limit in \eqref{CauchyFluxasLimit} does not exist for all smooth surfaces in $\Omega$~\cite{podio2006concentrated}, e.g., if $\boldsymbol{\sigma}=\delta_O \boldsymbol{I}$ then $\mathcal{F}$ is well defined only for surfaces $\mathcal{S}$ such that $O \notin \mathcal{S}$. The stress field $\boldsymbol{\sigma}\in \mathcal{D}'(\Omega,\Sym)$ can not be then interpreted in terms of the Cauchy flux map and \eqref{BalanceLaw} can no longer be used as the general equilibrium condition since $\mathcal{F}(\partial \mathcal{P})$ is not defined for arbitrary $\mathcal{P}$. The distributional stress equilibrium condition, which has no such restriction, can also be interpreted as the distributional limit of equilibrium of smooth stress fields. Indeed, for any stress field $\boldsymbol{\sigma}\in \mathcal{D}'(\Omega,\Sym)$ there exists a sequence of smooth maps $\boldsymbol{\sigma}_\rho$ such that $\boldsymbol{\sigma}_\rho \to \boldsymbol{\sigma}$ as $\rho \to 0$~\cite[Section~5.2]{friedlander1998introduction}. Analogously, the body force field $\boldsymbol{B}\in \mathcal{D}'(\Omega,\mathbb{R}^2)$ is interpreted as the limiting value of a sequence of smooth body force fields $\boldsymbol{B}_\rho$. We say that the distributional stress field is in equilibrium if it is the limit of a sequence of smooth equilibrated stress fields.  The equilibrium condition \eqref{DistributionalBalanceLaw} follows as the limit of the conditions $\div \boldsymbol{\sigma}_\rho + \boldsymbol{B}_\rho=\boldsymbol{0}$ in $\Omega$ as $\rho \to 0$. 

\section{Stress fields with interfacial concentration} \label{concent}
\begin{figure}
\begin{center} 
\includegraphics[scale=0.6]{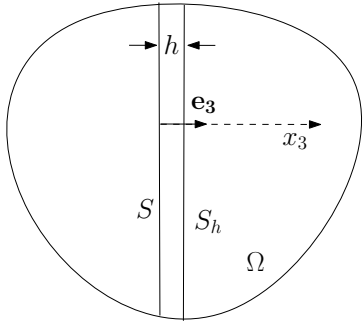}
\caption{Planar surfaces $S$ and $S_h$, both parallel to each other, separated by a distance $h$.}
\end{center}  \label{figdipole}
\end{figure}

We now derive local consequences of the distributional equilibrium condition \eqref{DistributionalBalanceLaw} taking the stress field to be of a specific form. We assume that the stress field $\boldsymbol{\Sigma} \in \mathcal{D}'(\Omega,\Sym)$ is such that $\boldsymbol{\Sigma}=\boldsymbol{\Sigma}_0+\boldsymbol{\Sigma}_1 +\boldsymbol{\Sigma}_2$, where $\boldsymbol{\Sigma}_0\in \mathcal{B}(\Omega,\Sym)$ with $\boldsymbol{\Sigma}_0(\boldsymbol{\psi})=\int_{\Omega}\langle\boldsymbol{\sigma},\boldsymbol{\psi}\rangle \dv$, $\boldsymbol{\Sigma}_1\in \mathcal{C}(\Omega,\Sym)$ with $\boldsymbol{\Sigma}_1(\boldsymbol{\psi})=\int_{S}\langle\boldsymbol{\sigma}_1,\boldsymbol{\psi}\rangle \da$, and $\boldsymbol{\Sigma}_2\in \mathcal{F}(\Omega,\Sym)$ with $\boldsymbol{\Sigma}_2(\boldsymbol{\psi})=\int_{S}\langle\boldsymbol{\sigma}_2,\partial \boldsymbol{\psi}/\partial {n}\rangle \da$, for all $\boldsymbol{\psi} \in \mathcal{D}(\Omega,\Sym)$. We consider the body force field $\boldsymbol{B}\in \mathcal{D}'(\Omega,\mathcal{V})$ such that $\boldsymbol{B}=\boldsymbol{B}_0+\boldsymbol{B}_1 +\boldsymbol{B}_2$, where $\boldsymbol{B}_0\in \mathcal{B}(\Omega,\mathcal{V})$ with $\boldsymbol{B}_0(\boldsymbol{\psi})=\int_{\Omega}\langle\boldsymbol{b},\boldsymbol{\psi}\rangle \dv$, $\boldsymbol{B}_1\in \mathcal{C}(\Omega,\mathcal{V})$ with $\boldsymbol{B}_1(\boldsymbol{\psi})=\int_{S}\langle\boldsymbol{b}_1,\boldsymbol{\psi}\rangle \da$, and $\boldsymbol{B}_2\in \mathcal{F}(\Omega,\mathcal{V})$ with $\boldsymbol{B}_2(\boldsymbol{\psi})=\int_{S}\langle\boldsymbol{b}_2,\partial \boldsymbol{\psi}/\partial {n}\rangle \da$, for all $\boldsymbol{\psi} \in \mathcal{D}(\Omega,\mathcal{V})$. In these definitions, $\boldsymbol{\sigma}$ and $\boldsymbol{b}$ are piecewise smooth (on $\Omega$) bulk Cauchy stress and body force fields; $\boldsymbol{\sigma}_1$ and $\boldsymbol{b}_1$ are smooth (on $S$) interfacial concentrations in stress (the interfacial Cauchy stress) and body force fields; $\boldsymbol{\sigma}_2$ and $\boldsymbol{b}_2$ are smooth (on $S$) interfacial concentrations in stress dipole and body force dipole fields. Both the bulk fields and the first set of interfacial concentration fields are well known~\cite{gurtin1973linear, gurtin1975continuum}. Recall that $\boldsymbol{\sigma}_1$ and $\boldsymbol{b}_1$  represent the excess fields (in the Gibbsian sense) representing contact and body forces not captured through their bulk counterparts.  To interpret $\boldsymbol{\Sigma}_2$, and hence the corresponding areal density $\boldsymbol{\sigma}_2$, as a surface stress dipole like term we consider a distribution $\boldsymbol{\Sigma}_h \in \mathcal{D}'(\Omega,\Sym)$ of the form
\begin{equation}
\boldsymbol{\Sigma}_h (\boldsymbol{\psi})=-\int_S \left\langle \frac{\boldsymbol{\sigma}_0}{h},\boldsymbol{\psi} \right\rangle\da + \int_{S_h} \left\langle \frac{\boldsymbol{\sigma}_0}{h},\boldsymbol{\psi} \right\rangle\da,
\end{equation}
for any $\boldsymbol{\psi}\in \mathcal{D}(\Omega,\Lin)$, where $\boldsymbol{\sigma}_0 \in \Sym$ is constant and $h$ is the separation distance between surfaces $S$ and $S_h$, both of which are assumed to be planar and parallel to each other (with normal along $\boldsymbol{e}_3$), see Figure~\ref{figdipole}. The distribution $\boldsymbol{\Sigma}_h$ represents two uniform stress concentrations, with opposite sign, on surfaces $S$ and $S_h$, both of which scale as the inverse of the distance between the planes. In the limit of an infinitesimal gap between the planes, i.e., $h \to 0,$ we obtain $\boldsymbol{\Sigma}_h \to \boldsymbol{\Sigma}_2$ where $\boldsymbol{\Sigma}_2(\boldsymbol{\psi})=\int_S \left\langle \boldsymbol{\sigma}_0,\partial \boldsymbol{\psi}/\partial {n} \right\rangle \da$, for any $\boldsymbol{\psi}\in \mathcal{D}(\Omega,\Lin)$. An analogous interpretation holds for $\boldsymbol{B}_2$.
For the considered form of  $\boldsymbol{\Sigma}$ and $\boldsymbol{B}$, and using Identities~\ref{DivergenceLemma}, it is straightforward to establish the equivalence of equilibrium condition \eqref{DistributionalBalanceLaw} with the following local conditions:\begin{subequations}
\label{EquilibriumConditionStrongForms}
\begin{align}
\div \boldsymbol{\sigma}+\boldsymbol{b}=\boldsymbol{0} ~\text{in}~\Omega-S,
\\
-\llbracket \boldsymbol{\sigma}\rrbracket\boldsymbol{n} + \div_S \boldsymbol{\sigma}_1 + \kappa \boldsymbol{\sigma}_1 \boldsymbol{n}-\div_S (\nabla_S \boldsymbol{n} \boldsymbol{\sigma}_2)+\boldsymbol{b}_1=\boldsymbol{0} ~\text{on}~S,
\\
-\boldsymbol{\sigma}_1 \boldsymbol{n}+\div_S \boldsymbol{\sigma}_2 +\boldsymbol{b}_2  = \boldsymbol{0} ~\text{on}~S, ~\text{and}
\\
\boldsymbol{\sigma}_2 \boldsymbol{n} = \boldsymbol{0} ~\text{on}~S.
\end{align}
\end{subequations}
When $\boldsymbol{\sigma}_2 = \boldsymbol{0}$ these conditions reduce to the well known stress equilibrium conditions (as mentioned in the first paragraph of Section~\ref{intro}), except that $\boldsymbol{\sigma}_1 \boldsymbol{n}= \boldsymbol{b}_2$. Therefore $\boldsymbol{\sigma}_1 \boldsymbol{n}$ will be non-zero if and only if there is a concentration of body force dipoles on $S$. On the other hand, if $\boldsymbol{\sigma}_2$ is allowed to be non-trivial then it can appear in response to both $\boldsymbol{b}_1$ and $\boldsymbol{b}_2$. The local equilibrium equations \eqref{EquilibriumConditionStrongForms} take a simple form if we assume the stresses to be of a dilatational form, i.e., $\boldsymbol{\sigma}=p \boldsymbol{I}$, $\boldsymbol{\sigma}_1=p_1 (\boldsymbol{I}-\boldsymbol{n}\otimes \boldsymbol{n})$, and $\boldsymbol{\sigma}_2=p_2 (\boldsymbol{I}-\boldsymbol{n}\otimes \boldsymbol{n})$, where $p$ is a piecewise smooth scalar field over $\Omega$ and $p_1, p_2$ are smooth fields over $S$.  We obtain 
\begin{subequations}
\label{EquilibriumConditionFluidform}
\begin{align}
\nabla p +\boldsymbol{b}=\boldsymbol{0} ~\text{in}~\Omega-S, \label{locstresseqb11}
\\
-\llbracket p \rrbracket\boldsymbol{n} + \nabla_S p
_1 + \kappa p_1 \boldsymbol{n} - p_2 \div_S (\nabla_S \boldsymbol{n} )-\nabla_S \boldsymbol{n} \nabla_S p_2 +\boldsymbol{b}_1=\boldsymbol{0} ~\text{on}~S,~\text{and} \label{locstresseqb12}
\\
\nabla_S p_2 + \kappa p_2 \boldsymbol{n} +\boldsymbol{b}_2  = \boldsymbol{0} ~\text{on}~S. \label{locstresseqb13}
\end{align}
\end{subequations}
Projecting Equations \eqref{locstresseqb12} and \eqref{locstresseqb13} onto the normal we obtain $\llbracket p \rrbracket =  \kappa p_1 + \langle \boldsymbol{b}_1, \boldsymbol{n} \rangle$ and  $ \kappa p_2 +\langle \boldsymbol{b}_2, \boldsymbol{n} \rangle  = 0$, respectively. Projecting these onto the tangent plane of $S$ yield  $\nabla_S p
_1  - p_2 \div_S (\nabla_S \boldsymbol{n} )-\nabla_S \boldsymbol{n} \nabla_S p_2 +(\boldsymbol{I}-\boldsymbol{n}\otimes \boldsymbol{n})\boldsymbol{b}_1=\boldsymbol{0}$ and $\nabla_S p_2 +(\boldsymbol{I}-\boldsymbol{n}\otimes \boldsymbol{n})\boldsymbol{b}_2  = \boldsymbol{0}$, respectively.

\section{Stress function for a non-contractible domain} \label{noncon}
Gurtin~\cite{gurtin1963generalization} established the necessary and sufficient conditions on a divergence-free smooth stress field $\boldsymbol{\sigma}\in C^\infty(\Omega,\Sym)$ for there to exist a smooth stress function $\boldsymbol{\phi}\in C^\infty(\Omega,\Sym)$ such that $\curl\curl \boldsymbol{\phi}=\boldsymbol{\sigma}$. If the domain $\Omega \subset \mathbb{R}^3$ has a boundary $\partial \Omega$ with  $k$ ($k>1$) mutually disjoint connected components then there are $2(k-1)$ global conditions which need to be additionally satisfied by the divergence-free stress. The following lemma extends this result in the sense of distributions which, in particular, allows us to work with non-smooth stress fields as will be illustrated later. 
\begin{lemma}
\label{ExistenceDistributionalStressPotential}
Given a stress field $\boldsymbol{\Sigma} \in \mathcal{D}'(\Omega,\Sym),$ there exists a stress function $\boldsymbol{\Phi} \in \mathcal{D}'(\Omega,\Sym)$ such that $\Curl \Curl \boldsymbol{\Phi}=\boldsymbol{\Sigma}$ if and only if 
\begin{subequations}
\label{DistributionaConditionExistenceStressPotential}
\begin{align}
\label{DCESP1}
\boldsymbol{\Sigma} (\boldsymbol{\psi}) = 0 ~~\text{and}~
\\
\label{DCESP2}
(\boldsymbol{x}\times \boldsymbol{\Sigma})^T (\boldsymbol{\psi})=0,
\end{align}
\end{subequations}
for all $\boldsymbol{\psi} \in \mathcal{D}(\Omega,\Lin)$ satisfying $\curl \boldsymbol{\psi}=\boldsymbol{0}$.
\begin{proof}
Assume that there exists $\boldsymbol{\Phi}\in \mathcal{D}'(\Omega,\Sym)$ such that $\Curl \Curl \boldsymbol{\Phi}=\boldsymbol{\Sigma}$. With this relation for $\boldsymbol{\Sigma}$  \eqref{DCESP1} is identically satisfied. In order to establish \eqref{DCESP2},  let $\boldsymbol{K}^T=\Curl \boldsymbol{\Phi}$ and note the identities
\begin{equation}
\Curl(\boldsymbol{x}\times \boldsymbol{K}^T)= (\boldsymbol{x}\times \boldsymbol{\Sigma})^T + \tr(\boldsymbol{K})\boldsymbol{I} - \boldsymbol{K}^T
\end{equation} 
and $\tr(\Curl \boldsymbol{\Phi})=0$. Henceforth we can write $(\boldsymbol{x}\times \boldsymbol{\Sigma})^T=\Curl (\boldsymbol{x}\times \boldsymbol{K}^T + \boldsymbol{\Phi})$, which immediately leads to \eqref{DCESP2}. 
On the other hand, we can use Lemma \ref{LemmaNSCurlFields} with \eqref{DCESP1} and \eqref{DCESP2} to imply the existence of $\boldsymbol{K}_1 \in \mathcal{D}'(\Omega,\Lin)$ and $\boldsymbol{F} \in \mathcal{D}'(\Omega,\Lin)$ such that $\boldsymbol{\Sigma}=\Curl {\boldsymbol{K}_1}^T$ and $(\boldsymbol{x}\times \boldsymbol{\Sigma})^T=\Curl \boldsymbol{F}.$ Let $\boldsymbol{F}_1=\boldsymbol{F}-(\boldsymbol{x}\times {\boldsymbol{K}_1}^T)$. Then $\Curl \boldsymbol{F}_1 = {\boldsymbol{K}_1}^T-\tr({\boldsymbol{K}_1})\boldsymbol{I}$. The distribution $\boldsymbol{\Phi}=\sym(\boldsymbol{F}_1)$ satisfies  $\Curl \Curl \boldsymbol{\Phi}=\boldsymbol{\Sigma}$.
\end{proof}
\end{lemma}

As an application of the lemma we recall the stress field $\boldsymbol{\Sigma} \in \mathcal{D}'(\Omega,\Sym)$, such that $\boldsymbol{\Sigma}=\boldsymbol{\Sigma}_0+\boldsymbol{\Sigma}_1 +\boldsymbol{\Sigma}_2$, as introduced in the beginning of Section~\ref{concent}. The necessary and sufficient conditions for the existence of a stress function $\boldsymbol{\Phi}\in \mathcal{D}'(\Omega,\Sym)$, satisfying $\Curl \Curl \boldsymbol{\Phi} = \boldsymbol{\Sigma}$, are given by \eqref{DCESP1} and \eqref{DCESP2}. Equation \eqref{DCESP1} is equivalent to the local conditions \eqref{EquilibriumConditionStrongForms} in addition to the global condition
\begin{equation}
\label{NonLocalConservation1}
\int_{\partial \Omega_i} \boldsymbol{\sigma} \boldsymbol{n}  \da + \int_{\partial S \cap \partial \Omega_i}  \boldsymbol{\sigma}_{1} \boldsymbol{\nu} \dl - \int_{\partial S \cap \partial \Omega_i} (\boldsymbol{\sigma}_{2} \nabla_S \boldsymbol{n}) \boldsymbol{\nu}   \dl  =\boldsymbol{0}. 
\end{equation}
Equation \eqref{DCESP2} is equivalent to
\begin{equation}
\label{NonLocalConservation2}
\int_{\partial \Omega_i} \boldsymbol{x}\times (\boldsymbol{\sigma} \boldsymbol{n})  \da  + \int_{\partial S \cap \partial \Omega_i}  \boldsymbol{x}\times(\boldsymbol{\sigma}_{1} \boldsymbol{\nu}) \dl - \int_{\partial S \cap \partial \Omega_i} \boldsymbol{x}\times ((\boldsymbol{\sigma}_{2} \nabla_S \boldsymbol{n}) \boldsymbol{\nu})  \dl  =\boldsymbol{0}.
\end{equation}
Therefore Equations \eqref{EquilibriumConditionStrongForms}, \eqref{NonLocalConservation1}, and \eqref{NonLocalConservation2} are the necessary and sufficient conditions for there to exist a stress function for the considered stress field.
Given $\boldsymbol{\Phi} \in \mathcal{B}(\Omega,\Sym)$ such that $\boldsymbol{\Phi}(\boldsymbol{\psi})=\int_\Omega \langle \boldsymbol{\phi},\boldsymbol{\psi} \rangle \dv$, for all $\boldsymbol{\psi}\in \mathcal{D}(\Omega,\Lin),$ the stress $\boldsymbol{\Sigma}\in\mathcal{D}'(\Omega,\Sym)$, given by $\boldsymbol{\Sigma}=\Curl\Curl \boldsymbol{\Phi}$, is of the form $\boldsymbol{\Sigma}=\boldsymbol{\Sigma}_0+\boldsymbol{\Sigma}_1+\boldsymbol{\Sigma}_2$, where $\boldsymbol{\Sigma}_0\in \mathcal{B}(\Omega,\Sym)$ with $\boldsymbol{\Sigma}_0(\boldsymbol{\psi})=\int_{\Omega}\langle\boldsymbol{\sigma},\boldsymbol{\psi}\rangle \dv$, $\boldsymbol{\Sigma}_1\in \mathcal{C}(\Omega,\Sym)$ with $\boldsymbol{\Sigma}_1(\boldsymbol{\psi})=\int_{S}\langle\boldsymbol{\sigma}_1,\boldsymbol{\psi}\rangle \da$, and $\boldsymbol{\Sigma}_2\in \mathcal{F}(\Omega,\Sym)$ with $\boldsymbol{\Sigma}_2(\boldsymbol{\psi})=\int_{S}\langle\boldsymbol{\sigma}_2,\partial \boldsymbol{\psi}/\partial {n}\rangle \da$, for all $\boldsymbol{\psi} \in \mathcal{D}(\Omega,\Sym)$, such that the densities $\boldsymbol{\sigma}$, $\boldsymbol{\sigma}_1$, and $\boldsymbol{\sigma}_2$ are given by
\begin{subequations}
\begin{align}
\boldsymbol{\sigma}=\curl\curl \boldsymbol{\phi} ~\text{in}~\Omega-S,
\\
\boldsymbol{\sigma}_1= (\llbracket \curl \boldsymbol{\phi}\rrbracket \times \boldsymbol{n})^T + \curl_S (\llbracket \boldsymbol{\phi} \rrbracket\times \boldsymbol{n})^T-\kappa ((\llbracket \boldsymbol{\phi} \rrbracket\times \boldsymbol{n})^T\times \boldsymbol{n})^T ~\text{on}~ S,~\text{and}
\\
\boldsymbol{\sigma}_2=((\llbracket \boldsymbol{\phi} \rrbracket\times \boldsymbol{n})^T\times \boldsymbol{n})^T ~\text{on}~S,
\end{align}
\end{subequations}
where the surface curl of a smooth tensor field, say $\boldsymbol{a} \in C^\infty(S,\Lin)$, is a smooth tensor field $\curl_S \boldsymbol{a} \in C^\infty(S,\Lin)$ such that, for any fixed $\boldsymbol{d} \in \mathcal{V}$, $\left(\curl_S \boldsymbol{a}\right)^T \boldsymbol{d} = \div_S \left(\boldsymbol{a} \times \boldsymbol{d}\right)$.
In other words a piecewise smooth stress function field over $\Omega$ yields both the bulk stress field and the interfacial stress fields $\boldsymbol{\sigma}_1$ and $\boldsymbol{\sigma}_2$.


\medskip
 
\bibliographystyle{plain}
\bibliography{ref}

  \end{document}